# Molecular theory for the phase equilibria and cluster distribution of associating fluids with small bond angles

Bennett D. Marshall and Walter G. Chapman.[1]
Department of Chemical and Biomolecular Engineering
Rice University
6100 S. Main
Houston, Texas  77005

## Abstract

We develop a new theory for associating fluids with multiple association sites. The theory accounts for small bond angle effects such as steric hindrance, ring formation and double bonding.  The theory is validated against monte carlo simulations for the case of a fluid of patchy colloid particles with three patches and is found to be very accurate. Once validated, the theory is applied to study the phase diagram of a fluid composed of three patch colloids. It is found that bond angle has a significant effect on the phase diagram and the very existence of a liquid – vapor transition.

---

[1] Author to whom correspondence should be addressed.  Tel: (1) 713.348.4900.  Fax: (1) 713.348.5478.  Email:  wgchap@rice.edu.

**I: Introduction**

In a recent publication[1] we extended Wertheim's thermodynamic perturbation theory[2,3] (TPT) to include the effect of bond angle $\alpha_{AB}$ in the equation of state for associating fluids with two association sites (an *A* association site and a *B* association site). It was found that various modes of association became dominant in various bond angle ranges. In strongly associating systems with large $\alpha_{AB}$ chains were the dominant type of associated cluster, for moderate $\alpha_{AB}$ rings became dominant, and for small $\alpha_{AB}$ double bonded molecules were overwhelmingly favored. The theory accounted for each association possibility and the effect of bond angle was included in each contribution. The theory was tested against monte carlo simulation data and found to be highly accurate.

Primitive models for association such as this have gained renewed interest in recent years with the advent of patchy colloids.[4] Patchy colloids contain some number of discrete attractive patches which results in anisotropic potentials between colloids. The specific properties of these colloids can be manipulated by varying the size, strength, surface location and number of attractive patches. This ability to manipulate these anisotropic potentials gives researchers the ability to program colloids to self – assemble into desired structures.[5] Examples of experimental realizations include the self assembly of two patch colloids into a Kagome lattice[6], and the work by Wang et *al.*[7] where controlled synthesis of patchy colloids with specific valence allowed for the self assembly into colloidal molecules. There has also been an extensive number of theoretical and simulation studies on the thermodynamics, phase behavior and self assembly of patchy colloid fluids.[8-21] Patchy colloids are typically modeled using Wertheim's first order perturbation theory (TPT1) and a potential model for conical association sites introduced by Bol[22] and later Chapman[23] that became widely used in the patchy colloid community after Kern and Frenkel[24] introduced this potential as a primitive model for patchy colloids.

As discussed extensively in our previous paper,[1] the application of TPT1 assumes there is no steric hindrance between association sites (patches), no cycles of association bonds, and finally there is no double bonding. These assumptions are valid for large bond angles, but as bond angle is decreased these effects must be accounted for. This was accomplished for the two patch case by combining and extending the resummed perturbation theory of Wertheim[3], a modification of the ring graph of Sear and Jackson[25] and a modification of the double bonded dimer graph of Sear and Jackson[26]. While shown to be very accurate, our previous theory[1] is restricted to the two patch case. However, to achieve phase equilibria and percolation in patchy colloid fluids, a minimum of three patches is required.[21]

In this work we wish to extend our previous theory to allow for small bond angle effects in colloids with more than two patches. Instead of tackling the more general case in which we allow for small bond angle effects between each pair of patches, we will restrict our analysis to the case where small bond angle effects are only accounted for a single pair of patches. This will allow for a tractable and logical extension of our previous work (we will consider the more difficult general case in a future paper).[1] We then validate the theory by comparison to monte carlo simulation data for the three patch case. Once validated, we show that bond angle has a huge effect on the liquid – vapor equilibria of three patch colloids. Throughout the paper we refer to associating molecules as colloids; however, the results in this paper are equally applicable as an equation of state for hydrogen bonding fluids.

**II: General theory**

In this section the theory for colloids of diameter $d$ with a set of patches $\chi = \{A, B, C...\}$ will be developed. There are a total of $n(\chi)$ patches on the colloid. The center of each pair of patches is separated by a bond angle $\alpha_{SP}$ and the size of the patches is controlled by the angle $\beta_c$ which defines the solid angle of the patch $2\pi(1 - \cos\beta_c)$. A diagram of this type of colloid can be found in Fig. 1 for the three patch case $\chi = \{A, B, C\}$. The potential of interaction between two colloids is given by the sum of a hard sphere potential $\phi_{HS}(r_{12})$ and orientation dependant attractive patchy potential

$$\phi(12) = \phi_{HS}(r_{12}) + \sum_{S \in \chi} \sum_{P \in \chi} \phi_{SP}(12) \tag{1}$$

The notation $(1) \equiv (\vec{r}_1, \Omega_1)$ represents the position $\vec{r}_1$ and orientation $\Omega_1$ of colloid 1 and $r_{12}$ is the distance between the colloids. Here we follow Bol[22] and Chapman et al.[27] who employed a potential for conical association sites

$$\phi_{SP}(12) = \begin{cases} -\varepsilon_{SP}, & r_{12} \leq r_c \text{ and } \beta_{S1} \leq \beta_c \text{ and } \beta_{P2} \leq \beta_c \\ 0 & \text{otherwise} \end{cases} \tag{2}$$

which states that if colloids 1 and 2 are within a distance $r_c$ of each other and each colloid is oriented such that the angles between the site orientation vectors and the vector connecting the two segments, $\beta_{S1}$ for colloid 1 and $\beta_{P2}$ for colloid 2, are both less than the critical angle $\beta_c$, the two sites are considered bonded and the energy of the system is decreased by a factor $\varepsilon_{SP}$. To ensure that each patch can only bond once we choose $r_c = 1.1\sigma$ and $\beta_c = 27°$. Kern and Frenkel[24] where the first to realize the potential given by Eqns. (1) – (2) provided an excellent description of the interactions between patchy colloids.

In Wertheim's theory each bonding state of a molecule (colloid) is assigned a number density. The density of colloids bonded at the set of patches $\alpha$ is given by $\rho_\alpha$. To aid in the reduction to irreducible graphs, Wertheim introduced the density parameters $\sigma_\gamma$

$$\sigma_\gamma = \sum_{\alpha \subset \gamma} \rho_\alpha \tag{3}$$

where the empty set $\alpha = \varnothing$ is included in the sum. Two notable cases of Eq. (3) are $\sigma_\chi = \rho$ and $\sigma_o = \rho_o$; where, $\rho$ is the total number density of colloids and $\rho_o$ is the density of colloids not bonded at any patch (monomer density).

In Wertheim's theory the change in free energy (over hard sphere) due to association is[2]

$$\frac{A^{AS}}{Vk_BT} = \sigma_\chi \ln\left(\frac{\sigma_o}{\sigma_\chi}\right) + Q + \sigma_\chi - \Delta c^{(o)}/V \tag{4}$$

where $V$ is the system volume, $T$ is temperature and $Q$ is given by

$$Q = -\sigma_\chi + \sum_{\substack{\gamma \subset \chi \\ \gamma \neq \varnothing}} c_\gamma \sigma_{\chi-\gamma} \tag{5}$$

The term $\Delta c^{(o)}$ is the associative contribution to the fundamental graph sum which encodes all anisotropic attractions between the colloids.

In the current work we wish to extend our theory for two patch colloids[1] with small bond angles to the case where there are more than two patches. In the two patch case it was shown that the bond angle dependence saturates near $\alpha_{AB} = 90°$; meaning for $\alpha_{AB} > 90°$ TPT1 is adequate. In order to make the derivation tractable and easily followed we will assume that the only small bond angle is $\alpha_{AB}$. This means that all interactions which do not involve patches A or B will be treated in TPT1. As in our previous work[1], we will assume that patches A and B do not self attract, that is $\varepsilon_{AA} = \varepsilon_{BB} = 0$.

We will split the fundamental graph sum into the TPT1 contribution $\Delta c_{TPT1}^{(o)}$ and the higher order contribution $\Delta c_{HOAB}^{(o)}$ which corrects for the small $\alpha_{AB}$.

$$\Delta c^{(o)} = \Delta c_{TPT1}^{(o)} + \Delta c_{HOAB}^{(o)} \tag{6}$$

The TPT1 contribution is given by[2]

$$\Delta c_{TPT1}^{(o)}/V = \frac{1}{2}\sum_{S \in \chi}\sum_{P \in \chi} \sigma_{\chi-S} \xi \kappa f_{SP} \sigma_{\chi-P} \tag{7}$$

The term $f_{SP} = \exp(\varepsilon_{SP}/k_B T) - 1$ is the magnitude of the association Mayer function and $\kappa = (1-\cos\beta_c)^2/4$ is the probability that patch S on colloid 1 and patch P on colloid 2 are oriented such that association can occur. The term $\xi = 4\pi \int_d^{r_c} r^2 g_{HS}(r) dr$ is the integral of the hard sphere reference pair correlation function over the bonding volume. Since the range of the interaction is short, we use the excellent approximation[23] that within the bonding volume $r^p g_{HS}(r) = d^p g_{HS}(d)$. The term $p$ depends on density and is given by[1] $p = 17.87\eta^2 + 2.47\eta$, where $\eta = \pi\rho d^3/6$ is the packing fraction. With this approximation of $g_{HS}(r)$ we obtain $\xi$ as

$$\xi = 4\pi d^3 g_{HS}(d)\left(\frac{(r_c/d)^{3-p}-1}{3-p}\right) \tag{8}$$

For the higher order $\Delta c_{HOAB}^{(o)}$ we have contributions which account for steric hindrance between patches A and B, $\Delta c_{ch:AB}^{(o)}$, cycles of n association bonds, $\Delta c_n^{cycle}$, and colloids double bonded at patches A and B, $\Delta c_d^{(o)}$

$$\Delta c_{HOAB}^{(o)} = \Delta c_{ch:AB}^{(o)} + \sum_{n=3}^{\infty} c_n^{cycle} + \Delta c_d^{(o)} \tag{9}$$

The types of associated clusters which contain paths of AB bonds are illustrated in Fig. 2. In Eq. (9) we account for the fact that association at patch A gives rise to steric hindrance to patch B reducing it's available bonding volume with the contribution $\Delta c_{ch:AB}^{(o)}$. We obtain this contribution through a resummed perturbation theory. The derivation is given in Appendix A. Here we quote the final result

$$\frac{\Delta c_{ch:AB}^{(o)}}{V} = \sum_{S\in\chi}\sum_{P\in\chi}\sigma_{\chi-S}\sigma_{\chi-P}\kappa f_{SA}\frac{f_{PB}}{f_{AB}}\xi(\lambda-1) \tag{10}$$

The term $\lambda$ is defined as $\lambda \equiv \left(1+(1-\Psi)\kappa f_{AB}\sigma_{\chi-AB}\xi\right)^{-1}$; where $\Psi$ is the blockage integral discussed in detail in our previous work[1]. When the angle $\alpha_{AB}$ is large enough that patches A and B are independent $\Psi \to 1$, which results in $\lambda \to 1$ and $\Delta c_{ch:AB}^{(o)} \to 0$. When bonding at patch A completely blocks another colloid from approaching to bond at patch B, the blockage integral vanishes $\Psi \to 0$. Results of numerical calculations of $\Psi$ can be found in Fig. 3.

For the two patch case[1] it was shown that rings of association bonds (cycles) played a crucial role for $40° \leq \alpha_{AB} \leq 90°$. The case here is much more complex since there are many more cycle forming possibilities. For instance, a cycle containing three colloids, of the type given in Fig. 1, could contain two colloids bonded at both patches *A* and *B* and one colloid bonded at patches *A* and *C*; another possibility would be a cycle with 3 colloids bonded at patches *A* and *B* etc.. To account for all cycle forming possibilities we would have to enumerate each possible cycle "composition" for each cycle size. This would be a doable, but tedious task. Instead, since we are assuming $\alpha_{AB}$ is the only small bond angle, we will only account for cycles formed from *AB* bonds only. For this case the contributions $\Delta c_n^{cycle}$ are obtained by a simple extension of the two patch case[1]

$$\Delta c_n^{cycle}/V = \frac{\left(f_{AB}\sigma_{\chi-AB}\hat{g}_{HS}K\right)^n}{nd^3}\Gamma^{(n)} \tag{11}$$

Where $K = 4\pi\sigma^2(r_c - \sigma)\kappa$ and $\hat{g}_{HS}$ is defined as

$$\hat{g}_{HS} = g_{HS}(d)\frac{2^p}{(r_c/d+1)^p} \tag{12}$$

The integrals $\Gamma^{(n)}$ are related to the number of configurations where *n* colloids are bonded in a cycle. The numerical results are correlated with a skewed Gaussian function

$$\Gamma^{(n)} = A_n \exp\left(-B_n(\alpha_{AB}-C_n)^2\right)\left(1+erf\left(-D_n(\alpha_{AB}-C_n)\right)\right) \tag{13}$$

where $\alpha_{AB}$ is in degrees and the constants $A_n$, $B_n$, $C_n$, $D_n$ depend on ring size $n$ and are given in the original publication.[1] The only difference between the $\Delta c_n^{cycle}$ in Eq. (11) and $\Delta c_n^{ring}$ in ref [1] is the substitution $\rho_o \to \sigma_{\chi-AB}$. It should be noted that $\Delta c_n^{cycle}$ accounts for cycles of $AB$ association bonds in clusters and is in no way limited to stand alone associated "rings" of colloids.

The last contribution to consider in Eq. (9) is for colloids which are double bonded at patches $A$ and $B$. In accordance with Eq. (2) two colloids can double bond only when there are overlapping patches $\alpha_{AB} - 2\beta_c < 0$. This overlap is represented by the orange patches with dashed outlines in Fig. 2. For two colloids to be double bonded the vector connecting the center of the two colloids must pass through the surface of overlap of the two patches $A$ and $B$ on each colloid. The contribution $\Delta c_d^{(o)}$ is similar to the two patch case[1] and is given by

$$\Delta c_d^{(o)}/V = \left(f_{AB}\sigma_{\chi-AB}\right)^2 \xi I_d / 2 \tag{14}$$

The constant $I_d$ is the probability that two colloids are oriented such that double bonding at patches $A$ and $B$ occurs and is given as[1]

$$I_d = \begin{cases} 0 & \text{for } \alpha_{AB} - 2\beta_c \geq 0 \\ S_{AB}^2 / 16\pi^2 & \text{otherwise} \end{cases} \tag{15}$$

Where $S_{AB}$ is the solid angle of the overlap of the patches $A$ and $B$ and is given by[28]

$$S_{AB} = 4\cos^{-1}\left(\frac{\sin\gamma_{AB}}{\sin\beta_c}\right) - 4\cos(\beta_c)\cos^{-1}\left(\frac{\tan\gamma_{AB}}{\tan\beta_c}\right) \tag{16}$$

where $\gamma_{AB}$ is obtained through the relation $\gamma_{AB} = \tan^{-1}\left((1-\cos\alpha_{AB})/\sin\alpha_{AB}\right)$. The only difference between Eq. (14) and the two patch case is the substitution $\rho_o \to \sigma_{\chi-AB}$. Equation (14) accounts for the double bonding of patches $A$ and $B$ in larger associated clusters as well as double bonded dimers.

Figure 3 displays the geometric quantities $I_d$, $\Psi$ and $\Gamma^{(n)}$ for $n = 3$ - 4. As expected $\Psi$ vanishes for small $\alpha_{AB}$ due to steric hindrance and becomes unity for large $\alpha_{AB}$ when association at patch $A$ no longer interferes with the ability of a third colloid to bond to patch $B$ (or vice versa). The cycle integrals $\Gamma^{(n)}$ are peaked around an optimum bond angle for cycle formation and the maximums of $\Gamma^{(n)}$ decrease and shift to larger bond angles as $n$ increases. The double bonding integral $I_d$, which represents the probability two colloids are oriented such that double bonding occurs, vanishes for $\alpha_{AB} > 54°$. In the limiting case $\alpha_{AB} = 0°$ the integral $I_d \to \kappa$.

Now that $\Delta c^{(o)}$ has been approximated, the densities of all bonding states of the colloids are calculated in a self consistent manner. In Wertheim's theory the densities are obtained from the following relation[2] (this represents free energy minimization)

$$\frac{\rho_\gamma}{\rho_o} = \sum_{P(\gamma)=\{\tau\}} \prod_\tau c_\tau \qquad (17)$$

where $P(\gamma)$ is the partition of the set $\gamma$ into non-empty subsets and $c_\tau$ are given by (for $\tau \neq \varnothing$)

$$c_\tau = \frac{\partial \Delta c^{(o)}/V}{\partial \sigma_{\chi-\tau}} \qquad (18)$$

For instance we have $\rho_{ABC} = \rho_o(c_{ABC} + c_{AB}c_C + c_{BC}c_A + c_{CA}c_B + c_A c_B c_C)$. From Eqns. (6) and (18) we see that $c_\beta = 0$ for $n(\beta) > 2$ and $c_{SP} = 0$ for $SP \neq AB$; this results in the following rule for the densities obtained from (17)

$$\frac{\rho_\gamma}{\rho_o} = \begin{cases} \prod_{S \in \gamma} c_S \left(1 + \frac{c_{AB}}{c_A c_B}\right) & \text{for} \quad AB \in \gamma \\ \prod_{S \in \gamma} c_S & \text{otherwise} \end{cases} \tag{19}$$

Equation (19) can be further simplified as

$$\frac{\rho_\gamma}{\rho_o} = \begin{cases} \frac{\rho_{AB}}{\rho_o} \prod_{S \in \gamma - AB} \frac{\rho_S}{\rho_o} & \text{for} \quad AB \in \gamma \\ \prod_{S \in \gamma} \frac{\rho_S}{\rho_o} & \text{otherwise} \end{cases} \tag{20}$$

Using the site operators of Wertheim[2], Eq. (20) can be rewritten as

$$\hat{\sigma}_\gamma = \begin{cases} \hat{\sigma}_{AB} \prod_{S \in \gamma - AB} \hat{\sigma}_S & \text{for} \quad AB \in \gamma \\ \prod_{S \in \gamma} \hat{\sigma}_S & \text{otherwise} \end{cases} \tag{21}$$

where $\hat{\sigma}_\gamma = \sigma_\gamma / \sigma_o$. For $\sigma_{\chi - AB}$ we obtain the simple relation from Eq. (21)

$$\hat{\sigma}_\chi = \hat{\sigma}_{AB} \hat{\sigma}_{\chi - AB} \tag{22}$$

Defining the fraction of colloids *not* bonded at both patches A and B, $X_{AB} = \sigma_{\chi-AB}/\sigma_\chi$ we obtain from Eqns. (17) and (22)

$$X_{AB} = \frac{1}{\hat{\sigma}_{AB}} = \frac{1}{c_{AB} + (1+c_A)(1+c_B)} \tag{23}$$

Equation (22) allows us to obtain $\sigma_{\chi-S}$ from Eq. (21) as

$$\hat{\sigma}_{\chi-S} = \begin{cases} \hat{\sigma}_\chi / \hat{\sigma}_S & \text{for } S \neq A \text{ or } B \\ \prod_{P \in \chi-S} \hat{\sigma}_P & \text{otherwise} \end{cases} \tag{24}$$

Combining (19) and (24) we get the fraction of colloids not bonded at patch S, $X_S = \sigma_{\chi-S}/\sigma_\chi$

$$X_S = \begin{cases} \dfrac{1}{1+c_S} & \text{for } S \neq A \text{ or } B \\ (1+c_L)X_{AB} & \text{otherwise} \end{cases} \tag{25}$$

In Eq. (25) when $S = A$, $L = B$ and when $S = B$, $L = A$. To obtain Eq. (25) we used the following relationship which was developed using Eq. (22)

$$\prod_{P \in \chi-A} \hat{\sigma}_P = \frac{\hat{\sigma}_\chi \prod_{P \in \chi-A} \hat{\sigma}_P}{\hat{\sigma}_{AB}\hat{\sigma}_{\chi-AB}} = \frac{\hat{\sigma}_\chi \prod_{P \in \chi-A} \hat{\sigma}_P}{\hat{\sigma}_{AB} \prod_{P \in \chi-AB} \hat{\sigma}_P} = \frac{\hat{\sigma}_\chi \hat{\sigma}_B}{\hat{\sigma}_{AB}} \tag{26}$$

The last density relation we need is for $\hat{\sigma}_\chi$, which we obtain using Eqns. (21) – (22)

$$\frac{1}{\hat{\sigma}_\chi} = X_o = \frac{X_{AB}}{X_A X_B} \prod_{S \in \chi} X_S \qquad (27)$$

Where $X_o$ is the fraction of colloids which are monomers. The condition that $c_{SP} = 0$ for $SP \neq AB$ allows for simplification of the $Q$ function obtained from Eq. (5) as

$$Q = -\sigma_\chi + \sum_{S \in \chi} c_S \sigma_{\Gamma-S} + c_{AB} \sigma_{\Gamma-AB} \qquad (28)$$

Which using the results above can be further simplified as

$$Q/\sigma_\chi = \sum_{S \in \chi}(1 - X_S) + \frac{X_A X_B}{X_{AB}} - 2 \qquad (29)$$

The $c_S$ are obtained from Eq (18) as

$$c_S = \sum_{P \in \chi} \Delta_{SP} X_P \left(1 + \left\{\frac{f_{SA}}{f_{SP}} \frac{f_{PB}}{f_{AB}} + \frac{f_{PA}}{f_{SP}} \frac{f_{SB}}{f_{AB}}\right\}(\lambda - 1)\right) \qquad (30)$$

where $\Delta_{SP} = \xi \kappa f_{SP} \sigma_\chi$. Likewise $c_{AB}$ is obtained as

$$c_{AB} = -(1-\Psi)\sum_{S \in \chi}\sum_{P \in \chi} X_S X_P \Delta_{SA} \Delta_{PB} \lambda^2 + \sigma_\chi X_{AB} f_{AB}^2 \xi I_d + \sum_{n=3}^{\infty} (f_{AB} \hat{g}_{HS} K)^n (\sigma_\chi X_{AB})^{n-1} d^{-3} \Gamma^{(n)} \qquad (31)$$

Comparing Eqns. (7), (9), (10) ,(25) and (30) we obtain

$$\frac{\Delta c_{TPT1}^{(o)} + \Delta c_{ch:AB}^{(o)}}{N} = \frac{1}{2}\sum_{S \in \chi}(1 - \chi_S) + \frac{X_A X_B}{X_{AB}} - 1 \qquad (32)$$

Where $N = \sigma_\chi V$ is the number of colloids. Combining (4), (11), (14), (29) and (32) we obtain the final form of the free energy

$$\frac{A^{AS}}{Nk_BT} = \sum_{S \in \chi}\left(\ln X_S - \frac{X_S}{2} + \frac{1}{2}\right) + \frac{\Delta A_{AB}}{Nk_BT} \qquad (33)$$

Where the summation gives the standard TPT1 free energy[27] and $\Delta A_{AB}$ represents the correction for the fact that sites $A$ and $B$ interact beyond first order and is given by

$$\frac{\Delta A_{AB}}{Nk_BT} = \ln\left(\frac{X_{AB}}{X_A X_B}\right) - \sum_{n=3}^{\infty} \frac{\Lambda_n}{n} - \frac{\Xi_d}{2} \qquad (34)$$

Where we have defined the quantities

$$\Lambda_n = (f_{AB} X_{AB} \hat{g}_{HS})^n K^n \sigma_\chi^{n-1} \Gamma^{(n)} d^{-3} \qquad (35)$$

and

$$\Xi_d = (f_{AB} X_{AB})^2 \sigma_\chi \xi I_d \qquad (36)$$

To evaluate Eq. (33) the fractions $X_{AB}$ and $X_S$ must be known. The fractions $X_S$ are obtained from Eqns. (25) and (30). Solving Eqns. (23) and (25) we obtain for $X_{AB}$

$$X_{AB} - X_{AB}^2 c_{AB} - X_A X_B = 0 \qquad (37)$$

In general, a total of $n(\chi)+1$ equations must be solved ($n(\chi)$ equations for $X_S$ given by Eq. 25 and Eq. (37) for $X_{AB}$) to obtain all fractions. The problem can be further simplified if symmetries among patches can be exploited.

Equation (37) concludes our analysis for colloids with a set of patches $\chi$. The internal energy and chemical potential are calculated in Appendix B. In the development of the theory we have assumed small bond angle effects only play a significant role for the bond angle $\alpha_{AB}$. In the following section we specialize the theory to the 3 patch case $\chi = \{A, B, C\}$.

## III: Specialization to the 3 patch case $\chi = \{A, B, C\}$

In this section we apply the results of section II to the 3 patch case $\chi = \{A, B, C\}$. This type of colloid is depicted in Fig. 1. From Eq. (30) we obtain the $c_K$'s as

$$c_A = \Delta_{AB} \lambda X_B + \Delta_{AC} \lambda X_C$$

$$c_B = \Delta_{AB} \lambda X_A + \Delta_{BC} \lambda X_C \tag{38}$$

$$c_C = (\Delta_{CA} X_A + \Delta_{CB} X_B) \lambda + X_C \tilde{\Delta}$$

Where the quantity $\tilde{\Delta}$ is given by

$$\tilde{\Delta} \equiv \Delta_{CC} + \Delta_{CA}(\lambda - 1)\frac{f_{CB}}{f_{AB}} + \Delta_{CB}(\lambda - 1)\frac{f_{CA}}{f_{AB}} \tag{39}$$

Now the required fractions can be determined through Eqns. (25) and (38) as

$$X_A = \frac{X_{AB}(1 + \Delta_{BC} \lambda X_C)}{1 - \Delta_{AB} \lambda X_{AB}}$$

$$X_B = \frac{X_{AB}(1 + \Delta_{AC} \lambda X_C)}{1 - \Delta_{AB} \lambda X_{AB}} \tag{40}$$

$$\frac{1}{X_C} = 1 + \Delta_{CA} X_A \lambda + \Delta_{CB} X_B \lambda + X_C \tilde{\Delta}$$

Combining Eqns. (37) and (40) gives a closed equation for $X_{AB}$. Once the fraction $X_{AB}$ is determined the remaining fractions can be calculated using Eqn. (40). To compare to simulations we will use the fractions of colloids bonded $i$ times $X_i$ ($i = 0-3$) which are obtained from Eqns. (17) and (27) as

$$X_o = X_{AB} X_C$$

$$X_1 = X_o(c_A + c_B + c_C) \tag{41}$$

$$X_2 = X_o(c_{AB} + c_A c_B + c_A c_C + c_B c_C)$$

$$X_3 = X_o(c_{AB} c_C + c_A c_B c_C)$$

Lastly, we will calculate the fraction of colloids which are bonded at both patches $A$ and $B$ in linear chains of $AB$ bonds, rings of $AB$ bonds and double bonds. The total density of colloids bonded at both patches $A$ and B, $\tilde{\rho}_{AB}$ contains contributions for colloids which are bonded at $A$ and $B$ only, as well as a contribution from colloids which are fully bonded. That is,

$$\tilde{\rho}_{AB} = \rho_{AB} + \rho_{ABC} \tag{42}$$

Using Eqns. (17) and (42) and defining the fraction $\tilde{\chi}_{AB} = \tilde{\rho}_{AB} / \sigma_\chi$ we obtain

$$\tilde{\chi}_{AB} = X_o(c_{AB} c_C + c_{AB} + c_A c_B + c_A c_B c_C) \tag{43}$$

We also know that $\tilde{\chi}_{AB}$ must satisfy the relation

$$\tilde{\chi}_{AB} = \tilde{\chi}_{AB}^{ch} + \tilde{\chi}_{AB}^{d} + \sum_{n=3}^{\infty} \tilde{\chi}_{AB}^{(n)} \tag{44}$$

where $\tilde{\chi}_{AB}^{ch}$ is the fraction of colloids bonded at both A and B which are in a chain of AB bonds, $\tilde{\chi}_{AB}^{d}$ is the fraction of colloids double bonded at A and B, and finally $\tilde{\chi}_{AB}^{(n)}$ is the fraction of colloids bonded at both A and B in a cycle of n AB bonds. These contributions are depicted pictorially in Fig. 2. Comparing Eqns. (31), (42) – (44) the following relations can be deduced

$$\frac{\tilde{\chi}_{AB}^{d}}{X_o} = \sigma_{\chi} X_{AB} f_{AB}^2 \xi I_d (1+c_C)$$

$$\frac{\tilde{\chi}_{AB}^{(n)}}{X_o} = (f_{AB} \hat{g}_{HS} K)^n (\sigma_{\chi} X_{AB})^{n-1} d^{-3} \Gamma^{(n)} (1+c_C) \tag{45}$$

$$\frac{\tilde{\chi}_{AB}^{ch}}{X_o} = \Psi \sum_{S \in \chi} \sum_{P \in \chi} X_S X_P \Delta_{SA} \Delta_{PB} \lambda^2 (1+c_C)$$

For the case of total steric hindrance between patches A and B, bonding at both patches in a chain of AB bonds to two other colloids becomes impossible resulting in $\Psi \to 0$. From Eq. (45) we see that for this case $\tilde{\chi}_{AB}^{ch} \to 0$, showing that the resummed perturbation theory was indeed successful.

**IV: Model and Simulation**

To validate the theory we perform new monte carlo simulations for three patch colloids of the type depicted in Fig. 1. In a spherical coordinate system ($\theta$ is the polar angle, $\phi$ is the azimuthal angle) the center of patch $C$ is located on the $z$ axis at $\theta = 0$, the center of patch $A$ is at $\phi = 0$ and $\theta = \pi - \alpha_{AB}/2$, and finally the center of patch $B$ is located at $\phi = \pi$ and $\theta = \pi - \alpha_{AB}/2$.

We consider 2 specific cases. In case I patch $C$ is of the same type as patch $A$; that is $\varepsilon_{CB} = \varepsilon_{AB}$ and $\varepsilon_{CA} = \varepsilon_{CC} = 0$. Case I could be a primitive model for a hydrogen bonding fluid with 2 hydrogen and 1 oxygen sites (or vice versa). Also this can be used as a model for patchy colloids. If the only attractions are due to association it seems unlikely this type of colloid could undergo a liquid vapor transition due to the fact that the $B$ type patches are limiting. To study the effect of bond angle on phase equilibria we consider another type, case II. In case II, the $C$ patch is attracted to all three patch types $\varepsilon_{CB} = \varepsilon_{CA} = \varepsilon_{CC} = \varepsilon_{AB}$.

To test the theory we perform *NVT* (constant *N*, *V* and *T*) and *NPT* (constant pressure *P*, *V* and *T*) monte carlo simulations. The colloids interact with the potential given by Eq. (1) with $r_c = 1.1\sigma$ and $\beta_c = 27°$. The simulations were allowed to equilibrate for $10^6$ cycles and averages were taken over another $10^6$ cycles. A cycle is defined as *N* attempted trial moves where a trial move is defined as an attempted relocation and reorientation of a colloid. For the *NPT* simulations a volume change was attempted once each cycle. In this work we used $N = 864$ colloids.

## V: Results

In this section we will compare theoretical and simulation predictions for three patch colloids. *NVT* simulations were performed for the bond angle range $0° \leq \alpha_{AB} \leq 90°$ at two states, $\eta = 0.1$, $\varepsilon^* = \varepsilon_{AB}/k_BT = 8$ and $\eta = 0.35$, $\varepsilon^* = 7.5$. Figure 4 shows the fractions of colloids bonded $k$ times for case I. For both states, the fractions remain relatively constant in the bond angle range $60° < \alpha_{AB} < 90°$ and oscillate in the range $45° < \alpha_{AB} < 60°$. For $\alpha_{AB} < 45°$ all fractions decrease as $\alpha_{AB}$ is decreased, with the exception of $X_2$ which increases rapidly as $\alpha_{AB}$ is decreased. Theory and simulation are in excellent agreement. Figure 5 gives these same fractions for the model case II. Overall, the bond angle dependence is stronger for this case with the fractions varying over the full range of $\alpha_{AB}$. In comparing case I (Fig. 4) and case II (Fig. 5) the most notable difference is the $\alpha_{AB}$ dependence of $X_3$ for $\alpha_{AB} < 45°$. In case I, $X_3$ decreases with decreasing $\alpha_{AB}$ in this region while for case II the opposite is true.

To shed light on this behavior, Fig. 6 gives the fractions of colloids bonded at both patches $A$ and $B$ in the various cluster types, see Fig. 2, for the state $\eta = 0.1$ and $\varepsilon^* = 8$. The fractions $\tilde{\chi}_{AB}^d$ were easily measured using *NVT* simulations, so we report these simulated fractions in addition to the theoretical predictions. In the region $65° < \alpha_{AB} < 120°$ chains of $AB$ bonds dominate. In the region $40° < \alpha_{AB} < 80°$ triatomic cycles contribute significantly, becoming the dominant contribution to $\tilde{\chi}_{AB}$ in the range $50° < \alpha_{AB} < 65°$. For bond angles $\alpha_{AB} < 54°$ double bonding becomes a possibility and increases rapidly with decreasing $\alpha_{AB}$. For bond angles $\alpha_{AB} < 50°$, $\tilde{\chi}_{AB}^d$ becomes the dominant contribution to $\tilde{\chi}_{AB}$. As $\alpha_{AB}$ becomes small, $\tilde{\chi}_{AB}^d$ approaches unity. For small $\alpha_{AB}$, double bonding is strongly favored due to the fact that you get the energetic benefit of forming a double bond for the same entropic penalty as a single bond. Theory and simulation are in excellent agreement. The insets of Fig. 6 give the fractions in 4-mer and 5-mer cycles of $AB$ bonds. As can be seen, these contributions (as well as for all larger $AB$ cycles) are small at these conditions. Both cases give similar results in Fig. 6; although, ring formation is slightly stronger for case I.

Now we can explain the difference in the $\alpha_{AB}$ dependence of $X_3$ for $\alpha_{AB} < 45°$ between cases I and II observed in Figs. 4 and 5. In this range, double bonding of patches $A$ and $B$ dominates. For case I, patch $C$ is of the same type as patch $A$ with $\varepsilon_{CA} = \varepsilon_{CC} = 0$. Since double bonding is favored at these small bond angles, the majority of $B$ patches are occupied in $AB$ double bonds, which means there are very few $B$ patches available to bond with $C$ patches. This results in a decrease in $X_3$ as $\alpha_{AB}$ is decreased and double bonding increases. The situation for case II is different. For this case, patch $C$ bonds to all three patch types. When $\tilde{\chi}_{AB}^d$ approaches unity at small $\alpha_{AB}$ the colloids can still bond three times by filling in with $CC$ bonds. For this reason we note the behavior for case II that decreasing $\alpha_{AB}$ results in an increase in $X_3$ for $\alpha_{AB} < 45°$. This is the opposite behavior observed in case I. A third case (not shown) in which $\varepsilon_{CB} = \varepsilon_{CA} = \varepsilon_{AB}$ and $\varepsilon_{CC} = 0$ shows similar bond angle dependence for $X_2$ and $X_3$ as case I.

To further test the accuracy of the theory, Figs. 7 and 8 compare theoretical predictions and $NPT$ simulation results of the compressibility factor $Z = P/\sigma_\chi k_B T$ for cases I and II respectively. For each case we consider bond angles $\alpha_{AB} = 20°$ and $45°$ at association energies $\varepsilon^* = 2,4,8$. Overall, the theory does a good job in predicting the temperature and bond angle dependence of the compressibility factor. To better access the effect of bond angle on $Z$ we plot $Z$ versus bond angle for the states $\eta = 0.1$, $\varepsilon^* = 8$ and $\eta = 0.35$, $\varepsilon^* = 7.5$ in Fig. 9. For bond angles $\alpha_{AB} > 100°$ there are essentially no cycles (or double bonds). It is in this region that the compressibility factor is a minimum. In the range $54° < \alpha_{AB} < 100°$ there is a steady increase in cycles as $\alpha_{AB}$ is decreased. When *AB* cycles are formed, longer chains of *AB* bonds must be broken, this results in an increase in the compressibility factor. For $\alpha_{AB} < 54°$ double bonding of patches *A* and *B* is possible and rapidly becomes the dominant contribution to $\tilde{\chi}_{AB}$. The formation of double bonds will break larger clusters of associated colloids resulting in an increase in the compressibility factor of the system. For this reason *Z* increases rapidly in this region becoming a maximum at $\alpha_{AB} = 0°$. The compressibility factor of case II is always lower than that of case I due to the increased amount of association. In the limit of strong association and $\alpha_{AB} = 0°$ a fluid of case I colloids will be composed of dimers while a fluid of case II colloids will be composed of longer linear chains.

Lastly we consider the effect of the bond angle $\alpha_{AB}$ on the phase diagram of case II colloids in Fig. 10. As can be seen, the phase diagram is strongly dependant on $\alpha_{AB}$. Comparing the two cases $\alpha_{AB} = 60°$ and $80°$, it is clear that cycle formation decreases both the critical temperature $T_c$ and critical density $\rho_c$. Lattice simulations[8] have also shown that cycle formation has a substantial influence on phase equilibria. Decreasing $\alpha_{AB}$ further to $\alpha_{AB} = 50°$, double bonding becomes significant which results in a further decrease in $T_c$ and $\rho_c$ as compared to the $\alpha_{AB} = 60°$ case. Decreasing $\alpha_{AB}$ below $50°$, results in a rapid increase in *AB* double bonds (Fig. 6). This rapid increase in $\tilde{\chi}_{AB}^d$ with decreasing $\alpha_{AB}$, results in a rapidly decreasing $T_c$ and $\rho_c$ as larger extended clusters must be sacrificed to accommodate double bonds. When double bonding becomes dominant, it is impossible to form a liquid phase due to the fact that a colloid which is double bonded can bond to a maximum of two colloids. Liquid – vapor phase equilibria in this case is impossible.[21]

## VI: Conclusions

We have extended our previous work[1] for the case of two patch associating fluids with small bond angles to the many patch case. Our model is restricted by the fact that we only account for small bond angle effects for the bond angle $\alpha_{AB}$. We have also assumed in the derivation that there is no self attraction between *A* and *B* patches (ie: $\varepsilon_{AA} = \varepsilon_{BB} = 0$). These assumptions allowed for a manageable derivation of the equation of state. The new equation of state was tested against monte carlo simulations for two cases of 3 patch colloids and was found to be accurate. Once validated the new theory was used to predict the effect of the bond angle $\alpha_{AB}$ on the phase diagram of three patch colloids with $\varepsilon_{CB} = \varepsilon_{CA} = \varepsilon_{CC} = \varepsilon_{AB}$. It was found that decreasing bond angle decreases both the critical temperature and density. Once $\alpha_{AB}$ is sufficiently small, and double bonding dominates, the phase transition is quenched and phase equilibria is no longer possible.

The extent to which the general case, where we allow all bond angles to be small, can be treated in TPT still needs to be addressed. When considering multiple small bond angles the possibility that steric hindrance between more than two patches simultaneously may need to be addressed. We will consider the general case in a future publication. Also, classical density functional theories (DFT) for associating fluids based on Wertheim's theory[17, 29-33] have proven to be a powerful tool for the description of associating fluids at interfaces. Extension of the current theory to inhomogeneous systems in the form of DFT could be used to study the effect of bond angle in inhomogeneous associating fluids.

**References:**

1. B. D. Marshall and W. G. Chapman, Physical Review E **87**, 052307 (2013).
2. M. Wertheim, Journal of Statistical Physics **42** (3), 459-476 (1986).
3. M. Wertheim, The Journal of Chemical Physics **87**, 7323 (1987).
4. E. Bianchi, R. Blaak and C. N. Likos, Phys. Chem. Chem. Phys. **13** (14), 6397-6410 (2011).
5. F. Romano and F. Sciortino, Nature materials **10** (3), 171-173 (2011).
6. Q. Chen, S. C. Bae and S. Granick, Nature **469** (7330), 381-384 (2011).
7. Y. Wang, Y. Wang, D. R. Breed, V. N. Manoharan, L. Feng, A. D. Hollingsworth, M. Weck and D. J. Pine, Nature **491** (7422), 51-55 (2012).
8. N. G. Almarza, Physical Review E **86** (3), 030101 (2012).
9. E. Bianchi, J. Largo, P. Tartaglia, E. Zaccarelli and F. Sciortino, Physical review letters **97** (16), 168301 (2006).
10. D. de las Heras, J. M. Tavares and M. M. T. da Gama, Soft Matter **7** (12), 5615-5626 (2011).
11. A. Giacometti, F. Lado, J. Largo, G. Pastore and F. Sciortino, Journal of Chemical Physics **131** (17), 174114 (2009).
12. A. Giacometti, F. Lado, J. Largo, G. Pastore and F. Sciortino, arXiv preprint arXiv:1004.1064 (2010).
13. N. Gnan, D. d. l. Heras, J. M. Tavares, M. M. T. d. Gama and F. Sciortino, The Journal of Chemical Physics **137** (8), 084704 (2012).
14. Y. Kalyuzhnyi, H. Docherty and P. Cummings, The Journal of Chemical Physics **133**, 044502 (2010).
15. Y. Kalyuzhnyi, H. Docherty and P. Cummings, The Journal of Chemical Physics **135**, 014501 (2011).
16. B. D. Marshall, D. Ballal and W. G. Chapman, The Journal of Chemical Physics **137** (10), 104909 (2012).
17. B. D. Marshall and W. G. Chapman, The Journal of Chemical Physics **138**, 044901 (2013).
18. F. Romano, E. Sanz, P. Tartaglia and F. Sciortino, Journal of Physics: Condensed Matter **24** (6), 064113 (2012).
19. J. Russo, J. Tavares, P. Teixeira, M. M. T. da Gama and F. Sciortino, The Journal of Chemical Physics **135**, 034501 (2011).
20. J. M. Tavares, L. Rovigatti and F. Sciortino, The Journal of Chemical Physics **137** (4), 044901 (2012).
21. J. M. Tavares, P. Teixeira and M. T. da Gama, Molecular Physics **107** (4-6), 453-466 (2009).
22. W. Bol, Molecular Physics **45**, 605 (1982).
23. W. G. Chapman, PhD. Thesis. 1988, Cornell University: Ithaca, NY.
24. N. Kern and D. Frenkel, The Journal of Chemical Physics **118**, 9882 (2003).
25. R. P. Sear and G. Jackson, Physical Review E **50** (1), 386-394 (1994).
26. R. P. Sear and G. Jackson, Molecular Physics **82** (5), 1033-1048 (1994).
27. W. G. Chapman, G. Jackson and K. E. Gubbins, Molecular Physics **65** (5), 1057-1079 (1988).
28. O. Mazonka, arXiv:1205.1396v1 (2012).

29. P. Bryk, S. Sokowski and O. Pizio, The Journal of Chemical Physics **125**, 024909 (2006).
30. A. Bymaster and W. Chapman, The Journal of Physical Chemistry B **114** (38), 12298 - 12307 (2010).
31. B. D. Marshall, A. J. García-Cuéllar and W. G. Chapman, The Journal of Chemical Physics **136**, 154103 (2012).
32. B. D. Marshall, A. J. García-Cuéllar and W. G. Chapman, Molecular Physics **110** (23) 2927–2939 (2012).
33. Y. Yu and J. Wu, The Journal of Chemical Physics **117**, 2368 (2002).

**Appendix A: Extension of RTPT1 to the multi patch case**

In this appendix $\Delta c_{ch:AB}^{(o)}$, Eq. (10), is derived in the framework of resummed perturbation theory. Our derivation will draw on that of Wertheim[3] for the 2 patch case, and that of Kalyuzhnyi et al.[14, 15] for the case of multiple bonding per patch. In general, resummed perturbation theories account for blocking effects for three body and higher interactions. In the current work, when the angle $\alpha_{AB}$ is small, association at patch $A$ can block a third colloid from approaching and associating to patch $B$ (or vice versa). Keeping true to the approximations of Section II, we will only account for blockage effects for colloids which are bonded at patches $A$ and $B$. For this case $\Delta c_{ch:AB}^{(o)}$ is approximated as the infinite series of chain diagrams

$$\frac{\Delta c_{ch:AB}^{(o)}}{V} = \sum_{S \in \chi} \sum_{P \in \chi} \sigma_{\chi-S} \sigma_{\chi-P} \sum_{n=1}^{\infty} \sigma_{\chi-AB}^{n} I_{n+1} \tag{A1}$$

Where the integrals $I_n$ are given by

$$I_n = \frac{1}{\tilde{\Omega}^n} \int f_{PB}(12) f_{AB}(23) \ldots f_{AB}(n-1,n) f_{AS}(n,n+1) G_{HS}(1\ldots n+1) d(2) \ldots d(n+1) \tag{A2}$$

Where $d(1) = d\vec{r}_1 d\Omega_1$ and $\tilde{\Omega} = 8\pi^2$. In first order resummed perturbation theory (RTPT1) the function $G_{HS}(1\ldots n)$ is approximated as[14]

$$G_{HS}(1\ldots n) = g_{HS}(r_{n-1,n}) \prod_{k=1}^{n-2} g_{HS}(r_{k,k+1}) f_{HS}(r_{k,k+2}) \tag{A3}$$

where $f_{HS}$ is the hard sphere reference system Mayer function given by

$$f_{HS}(r) = e_{HS}(r) - 1 = \begin{cases} 0 & r \geq d \\ -1 & r < d \end{cases} \tag{A4}$$

The association Mayer functions can be decomposed as

$$f_{SP}(12) = f_{SP}\theta_{SP}(12) \tag{A5}$$

Where $f_{SP} = \exp(\varepsilon_{SP}/k_B T) - 1$, and $\theta_{SP}(12)$ is given by

$$\theta_{SP}(12) = \begin{cases} 1 & r_{12} \leq r_c \text{ and } \beta_{S1} \leq \beta_c \text{ and } \beta_{P2} \leq \beta_c \\ 0 & \text{otherwise} \end{cases} \tag{A6}$$

Now $f_{AS}(n, n+1)$ can be rewritten as

$$f_{AS}(n, n+1) = \frac{f_{AS}}{f_{AB}} f_{AB}\theta_{AS}(n, n+1) \tag{A7}$$

Combing Eqns. (A2), (A5) - (A7) we rewrite Eq. (A1) as

$$\frac{\Delta c_{ch:AB}^{(o)}}{V} = \sum_{S \in \chi}\sum_{P \in \chi} \sigma_{\chi-S}\sigma_{\chi-P} \frac{f_{AS}f_{PB}}{f_{AB}^2} \sum_{n=0}^{\infty} \sigma_{\chi-AB}^n \hat{I}_{n+1} - \sum_{S \in \chi}\sum_{P \in \chi} \sigma_{\chi-S}\sigma_{\chi-P} \frac{f_{AS}f_{PB}}{f_{AB}^2} \hat{I}_1 \tag{A8}$$

where $\hat{I}_n$ is given by

$$\hat{I}_n = \frac{f_{AB}^n}{\tilde{\Omega}^n} \int \theta_{AB}(12)\theta_{AB}(23)...\theta_{AB}(n-1,n)\theta_{AB}(n,n+1) G_{HS}(1...n+1) d(2)...d(n+1) \tag{A9}$$

and we have added and subtracted the contribution due to $\hat{I}_1$. In Eq. (A9) we made the substitutions $\theta_{PB}(12) \to \theta_{AB}(12)$ and $\theta_{AS}(12) \to \theta_{AB}(12)$; this does not affect the value of the integral as long as all patches are restricted to be the same size. Now, the infinite sum in Eq. (A8) can be approximated as described in ref[14] to yield

$$\frac{\Delta c_{ch:AB}^{(o)}}{V} = \sum_{S \in \chi} \sum_{P \in \chi} \sigma_{\chi-S} \sigma_{\chi-P} \frac{f_{AS} f_{PB}}{f_{AB}} \kappa \xi \left( \frac{1}{\left(1 + (1-\Psi) \kappa f_{AB} \sigma_{\chi-AB} \xi\right)} - 1 \right) \qquad (A10)$$

Where $\Psi = 1 + \hat{I}_2 / \hat{I}_1^2$ is the blockage integral, which was evaluated in our previous paper.[1]

**Appendix B: Derivation of chemical potential and internal energy**

Deriving the chemical potential through the free energy given by Eq. (33) would be a laborious process. A much simpler method is as follows. From the graphical results of Wertheim[2] the association contribution to the chemical potential can be written as (note we are considering only the association contribution to $c_o$)

$$\frac{\mu^{AS}}{k_B T} = \ln X_o - c_o^{AS} \tag{B1}$$

The term $c_o^{AS}$ is found to be

$$c_o^{AS} = \frac{\eta}{2}\frac{\partial \ln \xi}{\partial \eta}\sum_{S\in\chi}\sum_{P\in\chi} X_S \xi \kappa f_{SP}\sigma_\chi X_P + \eta\frac{\partial \ln \xi}{\partial \eta}\sum_{S\in\chi}\sum_{P\in\chi}\sigma_\chi X_S X_P \kappa f_{SA}\frac{f_{PB}}{f_{AB}}\xi(\lambda-1) \\ -(1-\Psi)\eta\frac{\partial \ln \xi}{\partial \eta}\sum_{S\in\chi}\sum_{P\in\chi} X_S X_P X_{AB}\Delta_{SA}\Delta_{PB}\lambda^2 + \frac{\Xi_d}{2}\eta\frac{\partial \ln \xi}{\partial \eta} + \sum_{n=3}^{\infty}\Lambda_n \eta\frac{\partial \ln \hat{g}_{HS}}{\partial \eta} \tag{B2}$$

Using Eq. (30) we can simplify Eq. (B2) as

$$c_o^{AS} = \frac{\eta}{2}\frac{\partial \ln \xi}{\partial \eta}\sum_{S\in\chi} X_S c_S - (1-\Psi)\eta\frac{\partial \ln \xi}{\partial \eta}\sum_{S\in\chi}\sum_{P\in\chi} X_S X_P X_{AB}\Delta_{SA}\Delta_{PB}\lambda^2 + \frac{\Xi_d}{2}\eta\frac{\partial \ln \xi}{\partial \eta} + \sum_{n=3}^{\infty}\Lambda_n \eta\frac{\partial \ln \hat{g}_{HS}}{\partial \eta} \tag{B3}$$

Using Eq. (25) to eliminate $c_S$, and Eq. (27) to eliminate $X_o$, we obtain the final form of the chemical potential

$$\frac{\mu^{AS}}{k_B T} = \sum_{S\in\chi}\ln X_S - \frac{1}{2}\sum_{S\in\chi}(1-X_S)\eta\frac{\partial \ln \xi}{\partial \eta} + \frac{\Delta \mu_{AB}}{k_B T} \tag{B4}$$

In Eq. (B4) the first two terms represent the TPT1 contribution and $\Delta\mu_{AB}$ provides the higher order corrections for small $\alpha_{AB}$ and is given by

$$\frac{\Delta\mu_{AB}}{k_B T} = \ln\left(\frac{X_{AB}}{X_A X_B}\right) + \left(1 - \frac{X_A X_B}{X_{AB}} - \frac{\Xi_d}{2} + (1-\Psi)\sum_{S\in\chi}\sum_{P\in\chi} X_S X_P X_{AB} \Delta_{SA}\Delta_{PB}\lambda^2\right)\eta\frac{\partial \ln \xi}{\partial \eta} - \sum_{n=3}^{\infty}\Lambda_n \eta \frac{\partial \ln \hat{g}_{HS}}{\partial \eta} \quad (B5)$$

The excess internal energy is obtained as

$$\frac{E^{AS}}{N} = \frac{\partial}{\partial\beta}\frac{\beta A^{AS}}{N} = \left(1 - \Xi_d - \sum_{n=3}^{\infty}\Lambda_n\right)\frac{\partial \ln X_{AB}}{\partial \beta} - \left(\Xi_d + \sum_{n=3}^{\infty}\Lambda_n\right)\frac{\partial \ln f_{AB}}{\partial \beta} + \sum_{S\neq A,B}\frac{\partial X_S}{\partial \beta}\left(\frac{1}{X_S} - \frac{1}{2}\right) - \frac{1}{2}\frac{\partial X_A}{\partial \beta} - \frac{1}{2}\frac{\partial X_B}{\partial \beta} \quad (B6)$$

where $\beta = 1/k_B T$, and the derivative $\partial X_{AB}/\partial\beta$ is obtained from Eq. (37) as

$$\frac{\partial X_{AB}}{\partial \beta} = \frac{X_{AB}^2 \frac{\partial c_{AB}}{\partial \beta} + X_A \frac{\partial X_B}{\partial \beta} + X_B \frac{\partial X_A}{\partial \beta}}{1 - 2X_{AB} c_{AB}} \quad (B7)$$

**Figure captions:**

**Figure 1:** Diagram of patchy colloid model

**Figure 2:** Examples of associated clusters with paths of *AB* bonds. Black lines represent an *AB* bond. Here we show 4-mer cycles, but cycles of all sizes are accounted for. Orange area gives surface of overlap of the *A* and *B* patches

**Figure 3:** Numerical results[1] for the blockage integral $\Psi$ which accounts for the steric hindrance between patches *A* and *B*, the probability that two colloids are oriented such that double bonding can occur $I_d$, and the ring integrals $\Gamma^{(3)}$ and $\Gamma^{(4)}$ which are proportional to the number of ring states for 3-mer and 4-mer rings respectively

**Figure 4:** The fractions of colloids bonded *k* times $X_k$ for case I at $\eta = 0.1$ and $\varepsilon^* = 8$ (top) and $\eta = 0.35$ and $\varepsilon^* = 7.5$ (bottom). Curves give theory predictions and symbols give simulation results

**Figure 5:** Same as Fig. 4 except for case II

**Figure 6:** Fractions bonded at both patches A and B in the various cluster types, Eq. (45), at $\eta = 0.1$ and $\varepsilon^* = 8$ for case I (top) and case II (bottom). Curves give theory predictions and symbols give simulation values for $\tilde{\chi}_{AB}^d$. Insets give cycle fractions $\tilde{\chi}_{AB}^{(4)}$ and $\tilde{\chi}_{AB}^{(5)}$. Larger cycle sizes are non-zero but negligible and are not shown

**Figure: 7:** Compressibility factor for case I colloids at bond angles $\alpha_{AB} = 20°$ (top) and $\alpha_{AB} = 45°$ (bottom) at association energies $\varepsilon^* = 2$ (short dashed line - theory, triangles - simulation), $\varepsilon^* = 4$ (long dashed line - theory, circles - simulation) and $\varepsilon^* = 8$ (solid line - theory, squares - simulation)

**Figure 8:** Same as Fig. 7 except for case II

**Figure 9:** Compressibility factor versus $\alpha_{AB}$ at $\eta = 0.1$ and $\varepsilon^* = 8$ (top) and $\eta = 0.35$ and $\varepsilon^* = 7.5$ (bottom)

**Figure 10:** Phase diagrams $T^* = 1/\varepsilon^*$ versus $\rho^* = \rho d^3$ for case II colloids at various bond angles $\alpha_{AB}$

**Figure 1:**

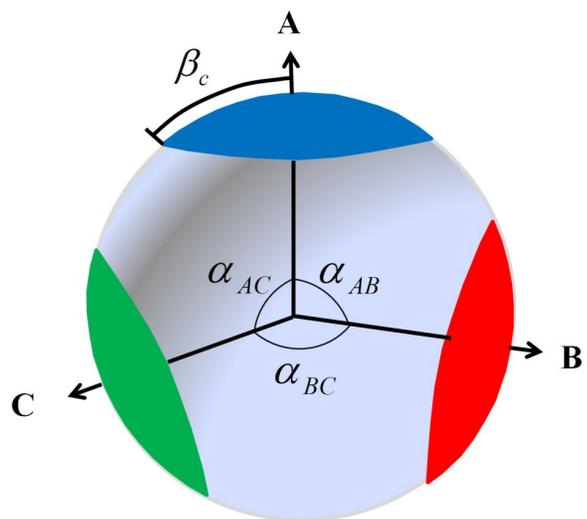

**Figure 2:**

**Chains of AB bonds**     **AB double bonds**

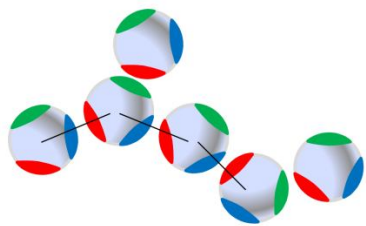
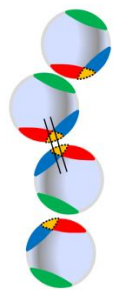

**Rings of AB bonds (cycles)**

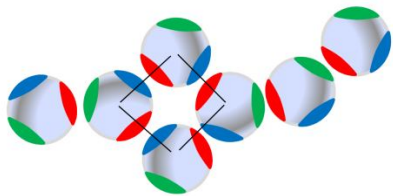

**Figure 3:**

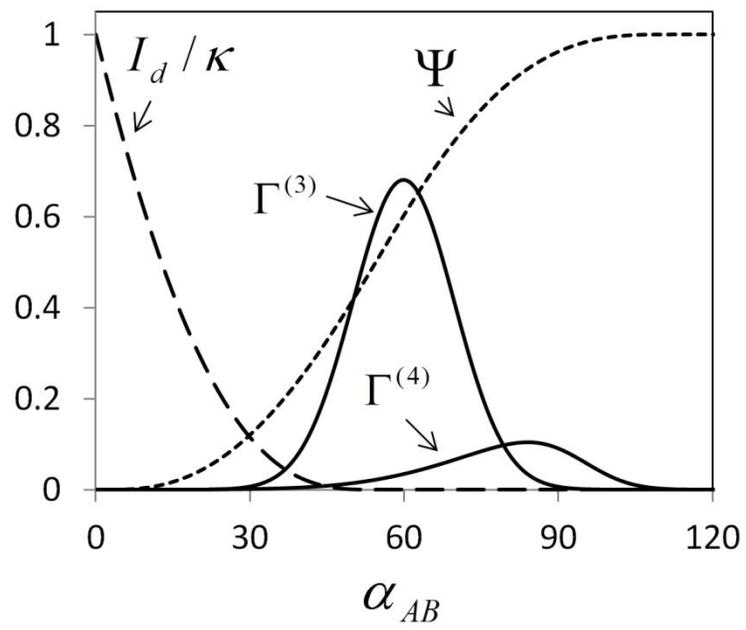

**Figure 4:**

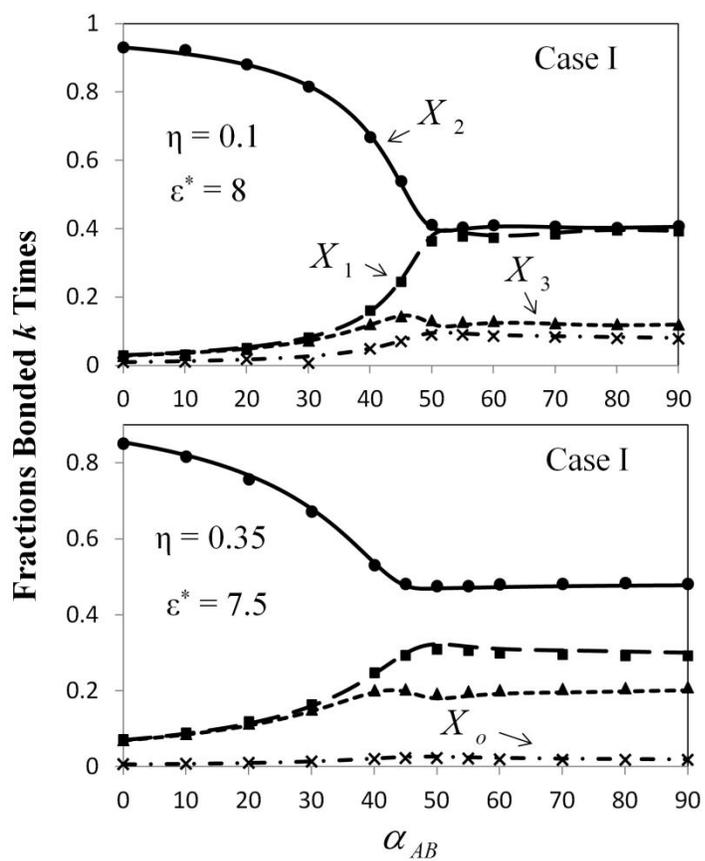

**Figure 5:**

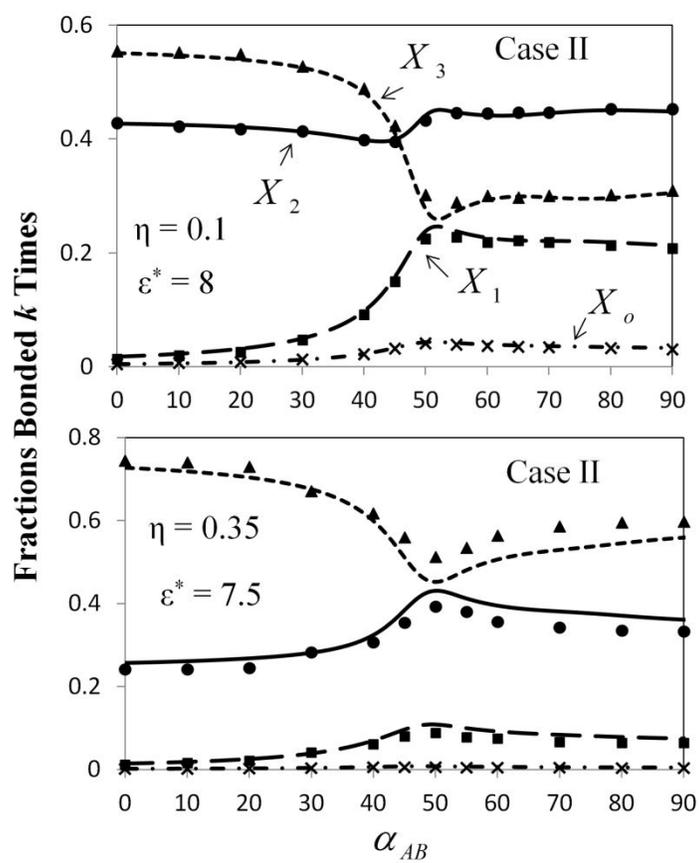

**Figure 6:**

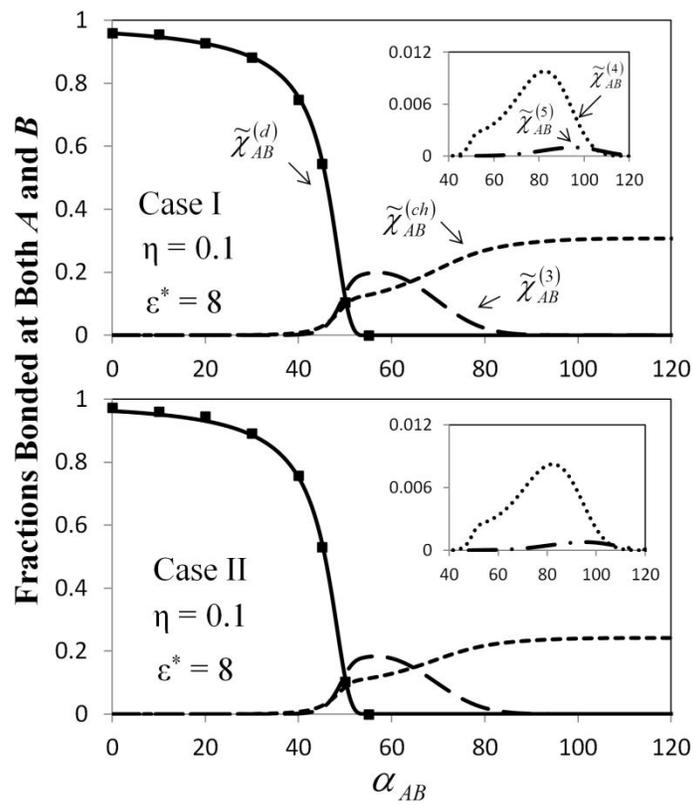

**Figure 7:**

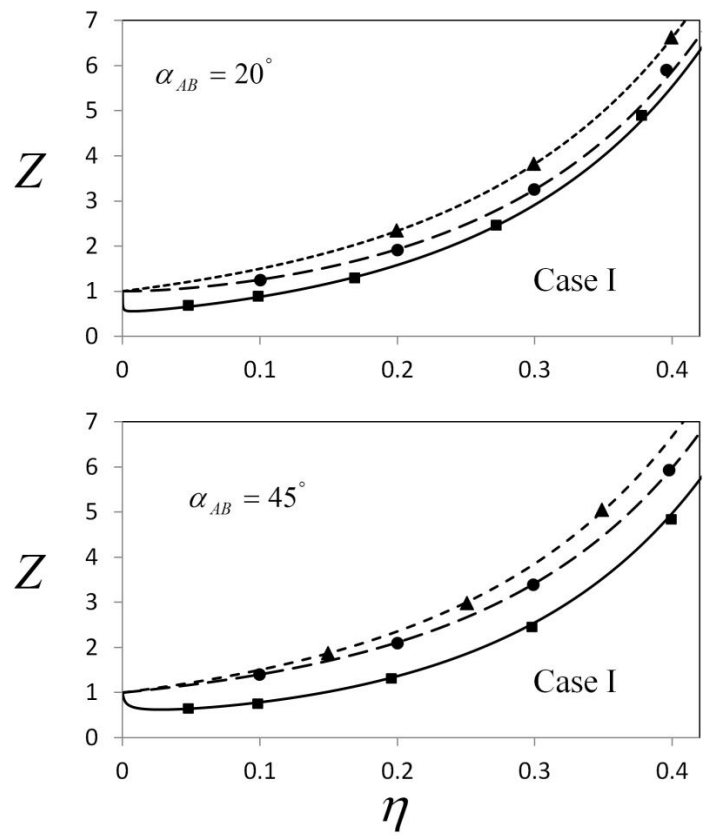

**Figure 8:**

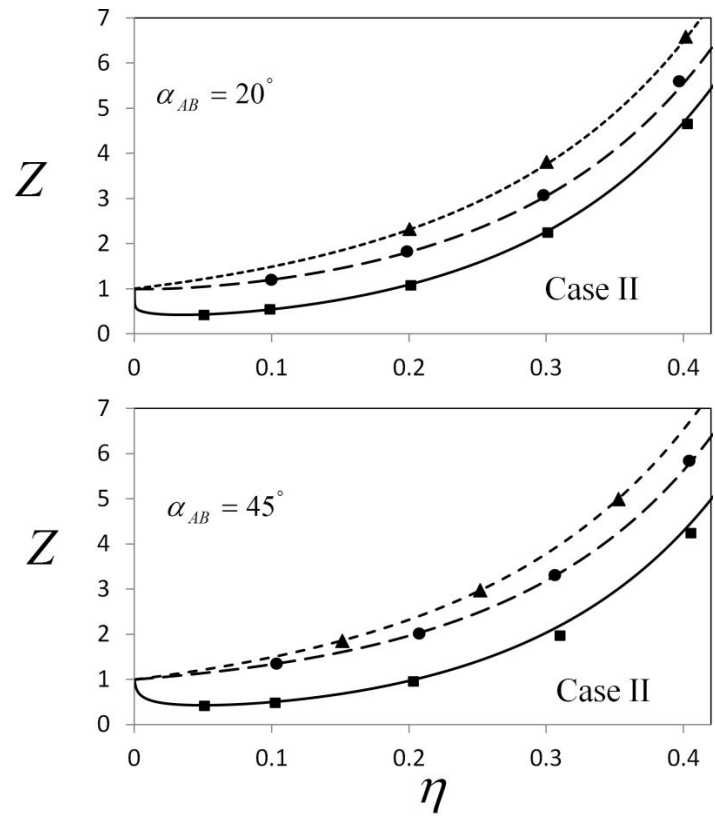

**Figure 9:**

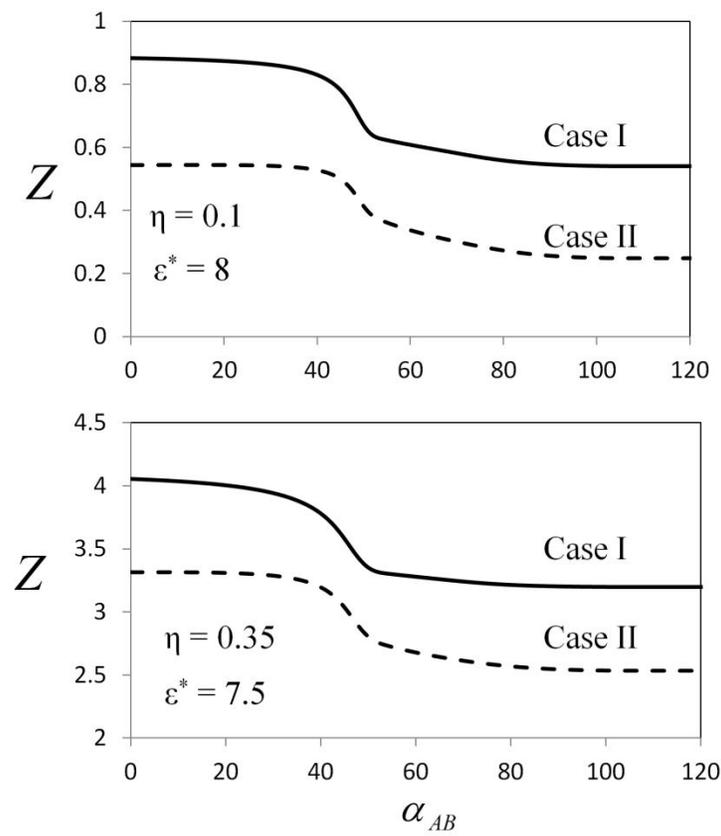

**Figure 10:**

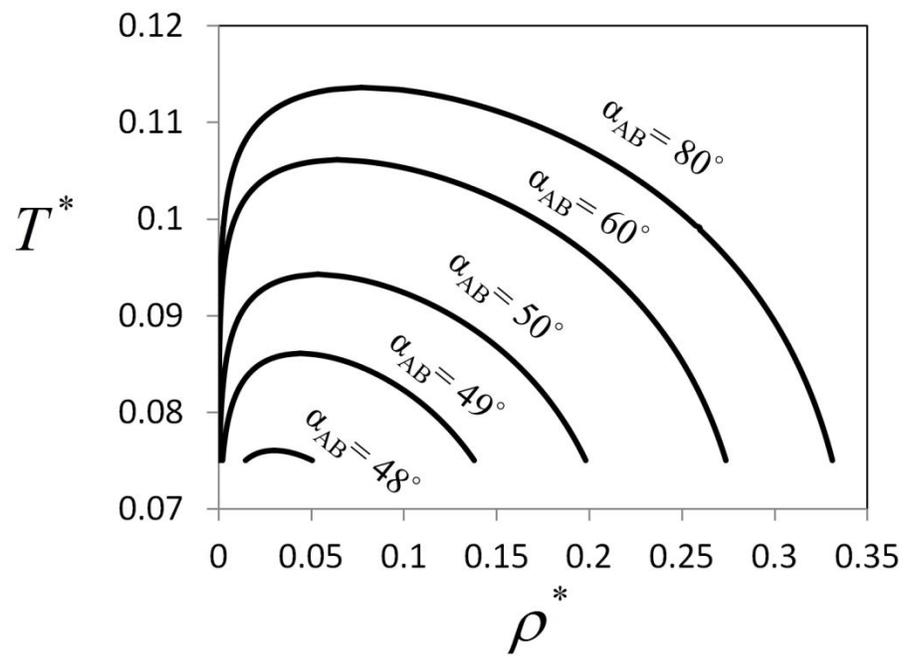